\documentclass[showpacs,preprintnumbers,amsmath,amssymb,superscriptaddress,unsortedaddress]{revtex4}
\usepackage{graphicx}% Include figure files
\usepackage{dcolumn}% Align table columns on decimal point
\usepackage{bm}% bold math
\usepackage{amsthm}
\usepackage{subfigure}

\newcommand{\bra}[1]{\langle #1 |}
\newcommand{\ket}[1]{| #1 \rangle}
\newcommand{\braket}[2]{\langle #1 | #2 \rangle}
\newcommand{\pro}[1]{\ket{#1}\bra{#1}}
\newcommand{\unidim}[2]{\ket{#1}\bra{#2}}
\def\beq{\begin{equation}}
\def\eeq{\end{equation}}
\def\id{{\rm id}}
\newcommand{\tr}{\operatorname{Tr}}

\newtheorem{theorem}{Theorem}
\newtheorem{lemma}{Lemma}

\newtheorem{definition}{Definition}
\newtheorem{remark}{Remark}

\def\be{\begin{equation}}
\def\ee{\end{equation}}
\def\ben{\begin{eqnarray}}
\def\een{\end{eqnarray}}
\def\bea{\begin{array}}
\def\eea{\end{array}}
\def\bei{\begin{itemize}}
\def\eei{\end{itemize}}

\def\ot{\otimes}

\renewcommand{\>}{\rangle}

\begin{document}

\preprint{}

\title{Entanglement-redistribution boxes}

\author{Andrzej Grudka}
\affiliation{Institute of Theoretical Physics and Astrophysics,
University of Gda\'{n}sk, 80-952 Gda\'{n}sk, Poland}

\affiliation{National Quantum Information Centre of Gda\'{n}sk, 81-824 Sopot, Poland}

\affiliation{Faculty of Physics, Adam Mickiewicz University, 61-614
Pozna\'{n}, Poland}

\author{Micha{\l} Horodecki}

\affiliation{Institute of Theoretical Physics and Astrophysics,
University of Gda\'{n}sk, 80-952 Gda\'{n}sk, Poland}

\affiliation{National Quantum Information Centre of Gda\'{n}sk, 81-824 Sopot, Poland}

\author{Pawe{\l} Horodecki}

\affiliation{National Quantum Information Centre of Gda\'{n}sk, 81-824 Sopot, Poland}

\affiliation{Faculty of Applied Physics and Mathematics, Technical
University of Gda\'{n}sk, 80-952 Gda\'{n}sk, Poland}

\author{Ryszard Horodecki}

\affiliation{Institute of Theoretical Physics and Astrophysics,
University of Gda\'{n}sk, 80-952 Gda\'{n}sk, Poland}

\affiliation{National Quantum Information Centre of Gda\'{n}sk, 81-824 Sopot, Poland}

\author{Marco Piani}

\affiliation{Institute of Theoretical Physics and Astrophysics,
University of Gda\'{n}sk, 80-952 Gda\'{n}sk, Poland}

\affiliation{Institute for Quantum Computing \& Department of Physics and Astronomy, University of Waterloo, Waterloo ON, Canada}

\date{\today}% It is always \today, today,

\begin{abstract}
We establish a framework to study the classical-communication properties of primitive local operations assisted by classical communication which realize various redistributions of entanglement, like, e.g., entanglement swapping.   On the one hand, we analyze what local operations and how much classical communication  are needed to perform them. On the other hand,  we investigate whether and to what extent such primitives can help to establish classical communication when they are used in the form of black boxes available to spatially-separated users. In particular, we find that entanglement swapping costs more communication than it can signal; in this sense, entanglement swapping is a weaker primitive than quantum teleportation.
\end{abstract}

\pacs{03.67.Lx, 42.50.Dv}% PACS, the Physics and Astronomy
\maketitle

\section{Introduction}
Entanglement~\cite{Horodecki}  is a purely quantum feature and a central resource in Quantum Information and Quantum Computation~\cite{Nielsen1}. Its interplay with classical communication has been the subject of many investigations. To name a few: entanglement together with classical communication enables quantum teleportation~\cite{Bennett5}; quantum communication assisted by entanglement provides dense coding~\cite{Bennett6}; classical communication allows to simulate stronger-than-classical correlations exhibited by entangled states~\cite{Toner1,Degorre1}. Furthermore, classical communication allows manipulation of entanglement in highly non-trivial ways, like in entanglement distillation~\cite{Bennett1} or entanglement swapping~\cite{Zukowski1}. In these two latter cases, classical communication can be considered a means to modify entanglement, in its ``quality'' or in its distribution among parties.

In this paper, in which we provide both detailed proofs of the claims presented in~\cite{Grudka2} and some new results, we contribute to the understanding of the interplay between entanglement and classical communication, by studying fundamental tasks, or primitives, dealing with \emph{entanglement redistribution}. In particular, we focus on the classical communication properties of multipartite operations that by means of Local Operations and Classical Communication (LOCC) perform said redistribution.

Recently, there has been a great progress in understanding classical communication properties of quantum bipartite operations (\cite {Bennett4,Childs1}). On the one hand, one estimates capacities of a bipartite operation: how many bits can one party send to another one, in the scenario where parties have access to many istances to the operation, which is thus considered a given resource? On the other hand, one investigated how much classical communication is needed to perform a given bipartite operation \cite{Beckman1}.

In this paper we concentrate on communication properties of primitives, irrespectively of their implementation. I.e., we assume that we deal with black-boxes which realize the task and we may not know the inner structure of the box, e.g., we do not know the specific unitaries or projectors involved in the redistribution process. Knowing only the task realized by the box, one can nonetheless ask how many bits can be sent by using the box. In contrast, one can ask what is the minimal number of bits necessary to realize a given primitive, i.e., for an optimal box.

\begin{figure}[!t]
\includegraphics[width=0.35\textwidth]{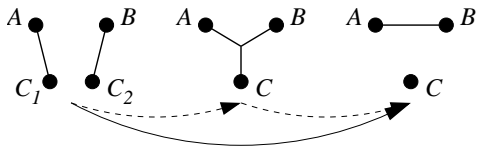}
\caption{Entanglement swapping (continuous arrow) in two steps (dashed arrows): from two EPR pairs to a GHZ state, to the final EPR pair. Lines denote quantum correlations.}
\label{fig:swappingsteps}
\end{figure}

We concentrate on entanglement redistribution-boxes, i.e., we deal with boxes which perform entanglement swapping (ES)~\cite{Bennett5,Zukowski1}, or create a GHZ state from two Bell states, or create a Bell state from a GHZ state (Figure \ref{fig:swappingsteps}). ES is a crucial ingredient of many protocols in quantum information theory. It is also one of the elements of quantum repeaters ~\cite{Briegel}, which enable long distance quantum communication.

On the one hand we define, the \emph{communication cost} (CC) of a \emph{particular} box,  i.e., one of the many possible boxes which realize a certain primitive, as the minimal amount of communication that is required to implement. On the other hand, we define the \emph{communication value} (CV) of the box as the minimal amount of communication that can be established with the box.

We further define the CC (CV) of a primitive as the minimal CC (CV) over all boxes realizing the primitive.

We will show that in general more communication has to be spent to perform a certain entanglement-redistribution, e.g., ES and creating a GHZ state from two Bell states, than can be established via any LOCC operation that realizes it. Hence, these two processes are irreversible with respect to classical communication. On the other hand we will show that creating a Bell state from a GHZ state is reversible with respect to classical communication.

The paper is organized as follows. In Section \ref{sec:definitions} we provide the definition of the basic concepts and quantities which we consider. In Section \ref{sec:main} we present our main results, which concern the structural characterization of entanglement-redistribution boxes, and upper and lower bounds on CC and CV of said oxes. Finally, we conclude in Section \ref{sec:conclusions}.

\section{Definitions}
\label{sec:definitions}

\subsection{Signalling and non-signalling quantum boxes}

When considering bipartite and multipartite operations, we can ask whether: (i) they need communication -- besides local operations and preshared correlations (entanglement) -- to be implemented~\footnote{When focusing on the necessity of communication to implement an operation, we may consider the communication resource to be only classical, as the quantum one may be obtained via teleportation}; (ii) they may be exploited in order to communicate. We respect to the latter question, it is clear that we can distinguish \emph{signaling} and \emph{non-signaling} (also known as \emph{causal}) quantum operations~\cite{Beckman1,Piani1}, as in Definition~\ref{def:signaling} below, where we consider the general multipartite case involving a set $\{A_1A_2\ldots A_n\}$ of $n$ parties. It is worth noticing that there are some operations on a bipartite system, that need communication to be performed, but anyway cannot be used to communicate~\cite{Beckman1,Piani1}.

\begin{definition}
\label{def:signaling}
An $n$-partite map $\Lambda\equiv\Lambda_{A_1A_2\ldots A_n}$ is \emph{signalling} from  $A_i$ to $A_j$ if there exist a state $\rho_{A_1A_2\ldots A_n}$ and two \emph{alphabet maps} $\Gamma^{k}\equiv\Gamma^{k}_{A_i}$, $k=0,1$, such that $\tr_{\backslash j}(\Lambda\circ\Gamma^{0}[\rho])\neq\tr_{\backslash j}(\Lambda\circ\Gamma^{1}[\rho])$. Otherwise, $\Lambda$ is said to be \emph{non-signalling} from $A_i$ to $A_j$.
\end{definition}

\begin{remark}
The concept of signalling (and non-signalling) map can be extended to the case where we consider some partition of the parties in disjoint sets  $\mathcal{P}_m=\{A_{i_p}\}_{p=1}^{|\mathcal{P}_m|}$, $\mathcal{P}_m\cap\mathcal{P}_n=\emptyset$ for $m\neq n$, $\bigcup\mathcal{P}_m = \{A_1A_2\ldots A_n\}$. Signalling and no-signalling from $\mathcal{P}_m$ to $\mathcal{P}_n$ are defined straightforwardly by substituting maps $\Gamma^{k}_{\mathcal{P}_m}$ and the operation of partial trace $\tr_{\backslash \mathcal{P}_n}$ in the definition above. Notice that, when we group parties, the alphabet maps as well as the ability of the receiving parties to ``decode the message'' may be restricted, e.g., to LOCC operations.
\end{remark}

Even though the possibility of signalling is defined in terms of an initial state which pertains to the same number of systems as the input of the map, it is possible to use further resources, i.e., a larger (in the number of systems) initial state of which only a part is subject to the map, in order to enhance communication. More specifically, it is possible to feed the map with only some subsystems of an entangled state.

\subsection{Communication cost and communication value}

In~\cite{Childs1} a general framework for analyzing classical communication properties for bipartite operations was studied. The scenario studied is the following: Alice and Bob have at disposal many uses of some bipartite operation $\Lambda$, and they want to take advantage of this fact to communicate classical messages one another, having possibly at disposal different amounts of pre-shared entanglement as a further resource. In particular, the authors considered the set of jointly achievable rates of classical communication from Alice to Bob and from Bob to Alice, and the rate of producing or consuming entanglement. 

Our analysis will differ in many ways from the one in~\cite{Childs1}. On the one hand, we will focus on a less general problem. First, we will consider only LOCC multipartite maps. Second, as suggested by the structure of the entanglement-redistribution boxes given by Theorem~\ref{thm:structure}, we will consider only one-way capacities of such multipartite LOCC maps, i.e., we will focus on optimizing only one of possibly many rates. On the other hand, we will also address the problem of the classical communication needed to realize LOCC maps.

In general, we will adopt two different points of view: that of the \emph{users} and that of the \emph{providers}. From the users' point of view, many copies of a bi- or multipartite operation are a resource to be consumed in order to achieve classical communication. From their point of view, the operation $\Lambda$ is a box, whose ``inner mechanism'' they might or might not  know, i.e., the box is a \emph{black-box}. Users feed the box with some input and they obtain an output from it. From the providers' point of view, whom we assume to be acting via LOCC, the operation is to be implemented, and it costs classical communication. Providers act ``inside'' the box, and must perform the map $\Lambda$ on any possible input from the users.

Since our focus is on LOCC operations to be exploited or implemented:
\begin{itemize}
\item we do allow users to consume  whatever amount of previously shared entanglement;
\item providers can not use entanglement in order to implement $\Lambda$: the only entanglement they deal with is the one fed to boxes by users. 
\end{itemize}

We are now ready to define the central quantities of our paper, which consist of rates and refer to the asymptotic case of many instances of boxes or tasks~\cite{Bennett4,Childs1}

\begin{definition}
The Communication Value (CV) of a box from the User $X_1$ to the User $X_2$ is the maximal rate (the capacity) at which User $X_1$ can reliably send classical messages to User $X_2$, i.e., the number of classical bits communicated per use of the box.
\end{definition}

\begin{definition}
The Communication Cost (CC) of a box from the Provider $X_1$ to the Provider $X_2$ is the rate at which classical communication must be sent from the first to the second in order to implement the box.
\end{definition}

\begin{definition}
The CC and CV from $X_1$ to $X_2$ for a task are the infima of CC and CV from $X_1$ to $X_2$ over all boxes realizing the task, i.e., the minimum amount of communication needed to realize the task, and the minimum amount of communication that we are sure it is possible to establish by means of any box that performs the task, respectively.
\end{definition}

As quantum mechanics respects causality, we have that CC -- both of a single box as well as of a task -- is always bigger than CV.

Because of the characterization of all ES-boxes we will provide, the CC and the CV will be analyzed with respect to the relevant direction $C\rightarrow AB$. If not differently stated, we will consider Alice and Bob as one party when considering the CV. This means that not only we allow communication between Alice and Bob, when they decode a message sent by Charlie, but we furthermore allow to perform global $AB$ quantum operations. Indeed, while the box is assumed to be (in its implementation) tripartite LOCC, this does not force the users to have the same limitation. Alternatively, one could think of pre-shared entanglement between Alice and Bob being consumed while trying to get a signal from Charlie by means of the box.

\subsection{Entanglement-redistribution black boxes}

The tasks we will study correspond to the redistribution of entanglement by means of LOCC operations. The entangled states relevant for our discussion are bipartite and tripartite: the maximal entangled EPR state of two qubits
\beq
\Psi^+_{X_1X_2}=\pro{\Psi^+}_{X_1X_2},\quad\ket{\Psi^+}_{X_1X_2}=\frac{1}{\sqrt{2}}(\ket{0}_{X_1}\ket{0}_{X_2}+\ket{1}_{X_1}\ket{1}_{X_2}) 
\eeq
and the GHZ state of three qubits
\beq
\ket{GHZ}_{X_1X_2X_3}=\frac{1}{\sqrt{2}}(\ket{0}_{X_1}\ket{0}_{X_2}\ket{0}_{X_3}+\ket{1}_{X_1}\ket{1}_{X_2}\ket{1}_{X_3}).
\eeq
We will consider multipartite LOCC operations. It is well known that such operations can be written in the form of separable operations, e.g., in the tripartite case we will consider, as:
\beq
\Lambda_{ABC}[\rho_{ABC}]=\sum_i A_iB_iC_i\rho_{ABC} A^\dagger_iB^\dagger_iC^\dagger_i,\quad \sum_i (A^\dagger_iB^\dagger_iC^\dagger_i)(A_iB_iC_i)=\openone,
\eeq
where we used the shorthand notation $A_iB_iC_i\equiv A_i\otimes B_i\otimes C_i$.

In particular, we will focus on operations which realize the three processes of redistribution of entanglement represented schematically in Figure~\ref{fig:swappingsteps}:
\begin{description}
\item[{[ES]}] Entanglement swapping: $\ket{\Psi^+}_{AC_1}\ket{\Psi^+}_{BC_2}\rightarrow \ket{\Psi^+}_{AB}$
\item[{[2EPR-GHZ]}] Two EPR pairs into a GHZ state: $\ket{\Psi^+}_{AC_1}\ket{\Psi^+}_{BC_2}\rightarrow \ket{GHZ}_{ABC}$
\item[{[GHZ-EPR]}] A GHZ state into an EPR pair: $\ket{GHZ}_{ABC}\rightarrow \ket{\Psi^+}_{AB}$
\end{description}
Thus, ES may be realized in two steps: 2EPR-GHZ first, followed by GHZ-EPR.

Both for [ES] and [GHZ-EPR] we refer to the $AB$ output alone, disregarding the output system of Charlie, i.e. effectively tracing it out.

An LOCC operation is said to be a box for one of the entanglement-redistribution tasks listed above, e.g., ES, if, given the correct input, e.g., two EPR pairs between the Users $A:C_1$ and $B:C_2$ in the case of ES, it outputs the correct output, e.g., an EPR pair between $A:B$; that is, it realizes the redistribution.

\begin{figure}
\subfigure[~Defining action of an ES-box. Time goes from left to right.]{\label{fig:ESbb}\includegraphics{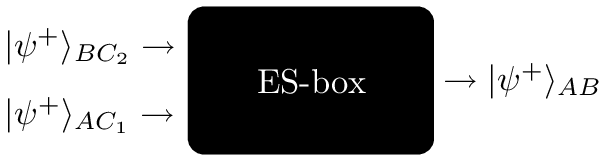}}\hspace{0.5cm}
\subfigure[~Actual (LOCC) implementation of a given ES-box.]{\label{fig:ESb-inside}\includegraphics{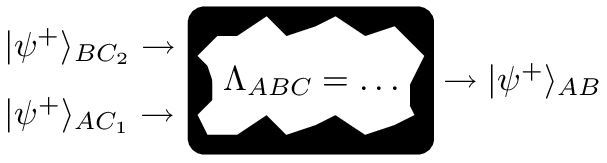}}
\caption{Entanglement swapping (black) box.} 
\end{figure}

The attribute of being ``black'' means that, e.g., users do not care about the inner implementation of the box, as long as it performs the corresponding task upon proper input (see Figure~\ref{fig:ESbb}). On the opposite side, providers have to deal with said inner implementation (see Figure~\ref{fig:ESb-inside}) by definition.

\section{Main results}
\label{sec:main}

We obtain two main results. First, we characterize completely the inner structure of entanglement-redistribution boxes. Second, using said characterization, we are able to compute bounds for CC and CV of both boxes and tasks, showing in particular a fundamental irreversibility regarding classical communication for entanglement swapping: entanglement swapping costs more classical communication than it can signal.

As an introduction to our main results, let us make some immediate observations about the communication properties of entanglement swapping.

We notice that any ES box is {\it signaling}
with respect to $C \rightarrow AB$.   Let us choose 
$|\Psi^{+}\rangle_{AC_{1}}|\Psi^{+}\rangle_{BC_{2}}$ as the initial state
and  the identity and the measurement
in the computational basis as Charlie's two alphabet maps. If Charlie applies
the identity then Alice and Bob obtain $|\Psi^{+}_{AB}\rangle$ as the output of the box.
If Charlie performs the measurement in computational basis, then he
destroys all the entanglement between the parties and Alice and Bob obtain a separable state as the output of the box.

Let us consider a specific example of ES-box. It is well known that quantum teleportation in particular realizes entanglement swapping. Hence, as our box we take \emph{any} LOCC teleportation protocol~\cite{Bennett5} from Charlie to Alice. CV of this box is equal to $2$ ~\cite{Bennett6}. In order to signal two bits from Charlie to Alice the parties take  $\ket{\Psi^+}_{AC_1}\ket{\psi}_B\ket{\Psi^+}_{A'C_2}$ as the initial state where $A'$ is local ancilla of Alice that is maximally entangled with $C_2$. Charlie can rotate locally the pair $\ket{\Psi^+}_{A'C_2}$ into one of four orthogonal Bell state. After the box is applied this Bell state is held only by Alice.  If we choose the standard teleportation protocol, constiting of a Bell measurement at the sender followed by a proper unitary rotation at the receiver, the CC is 2. For any other LOCC teleportation protocols,  CC  is greater or equal to 2, as cannot be less than CV because of causuality~\cite{Bennett5}. 

Interestingly, we will show that there are ES boxes which have CV
equal to $1$. We will also show that every
ES-box has CC greater or equal to $2$. Thus, there are ES-boxes which exhibit
communication irreversibility.

\subsection{Structure of boxes}

The first of our main results is a complete characterization of the structure of all entanglement-redistribution boxes. We prove that all the kinds of entanglement redistribution we are analyzing, ES, 2EPR-GHZ and GHZ-EPR, are essentially one-way protocols: the provider Charlie measures his part of the system, and communicates the result of the measurement to Alice and/or Bob, who then apply a rotation to their systems accordingly to the result they have been communicated.

\begin{theorem}
\label{thm:structure}
Any entanglement redistribution box $\Lambda_{ABC}$ is of the form $\Lambda(\rho_{ABC})={\rm Tr}_{C}\big(\sum_{i}U_{A}^{i}U_{B}^{i}E_{C}^{i}(\rho_{ABC})U_{A}^{i\dagger}U_{B}^{i\dagger}E_{C}^{i\dagger}\big)$,
where $U_{A}^{i}$ and $U_{B}^{i}$ are unitary operations, and $E_{C}^{i}$ depend on the kind of box in the following way:
\begin{description}
\item[{[ES]}] $E_{C}^{i}=\unidim{u^i}{\psi^i_+}$ are rank-one measurement operators, with $\ket{\psi^i_+}$ normalized maximally entangled states of Charlie's
particles, which satisfy
$\sum_{i}E_{C}^{i\dagger}E_{C}^{i}=\sum_i\|u_i\|^2\pro{\psi^i_+}={\openone}_{C_1C_2}$ for ES box;
\item[{[2EPR-GHZ]}] $E_{C}^{i}=c_i(\ket{c_0}\bra{a_0^ib_0^i}+\ket{c_1}\bra{a_1^ib_1^i})$ are rank-two measurement operators, with $\braket{c_m}{c_n}=\delta_{mn}$, and $\braket{a_r^i}{a_s^i}=\braket{b_r^i}{b_s^i}=\delta_{rs}$ for all $i$, satisfying $\sum_i|c_i|^2(\pro{a_0^ib_0^i}+\pro{a_1^ib_1^i})={\openone}_C$;
\item[{[GHZ-EPR]}] $E_{C}^{i}=\unidim{u^i}{\psi^i}$ are rank-one measurement operators, with $\ket{\psi^i}=\frac{1}{\sqrt{2}}(|0\rangle+e^{i\phi}|1\rangle)$, which satisfy 
$\sum_{i}E_{C}^{i\dagger}E_{C}^{i}=\sum_i\|u_i\|^2\pro{\psi^i}={\openone}_{C}$.
\end{description}
\end{theorem}

\begin{proof} The proof consists of two parts. In the first part
\emph{(I)} we prove that Alice and Bob cannot perform non-unitary
operations. In the second part \emph{(II)} we find the conditions
which have to be satisfied by Charlie's operators.

\emph{(I)} We will prove this part by contradiction. Every LOCC protocol corresponds to rounds of local operations followed by classical communication. Let us suppose that Alice is the first, between her and Bob, to perform an operation different from an isometry, and that this happens at round $n_0$~\footnote{Since isometric operations are deterministic, it actually is $n_0=1$ or $n_0=2$, i.e. either Alice acts first or after Charlie.}. After such an operation by Alice, we allow Bob and Charlie to join, so that the protocol may continue as an $A|BC$ \emph{bipartite} LOCC protocol.  If something is impossible even allowing this, then it is a fortiori impossible by continuing with a \emph{tripartite} LOCC protocol, as it actually is.

We may think of one step of any LOCC protocol as of the
application by one party of trace preserving
maps, which (may) depend on the result $i$ of the previous steps, of the form
$\Lambda_i(\rho)=\sum_{j}A_{ij} \rho A_{ij}^{\dag} \otimes | j
\rangle \langle j|$,
where $| j \rangle \langle j |$ are
classical registers -- locally available to all the parties -- indicating which Kraus operator is applied, and $A_{ij}$ are Kraus operators that may depend on the previous
results, i.e. on the previous state $i$ of some classical registers. Therefore, we can describe the evolution of the branches of the protocol, with an initial pure state maintaining its purity in each branch, and we can consistently analyze the intermediate steps of the protocol. The relevant, total channel is obtained by tracing out the classical registers.

Before the action of Alice, the state $\rho_{ABCR}$ of the system (the relevant system plus the classical registers) can be described as $\rho_{ABCR}=\sum_{i}p_{i}\rho^i_{A|BC}\otimes
|i \rangle \langle i |$, where $\rho^i_{A|BC}=|\Phi^{i}\rangle_{A|BC}\langle\Phi^{i}|_{A|BC}$, $\sum_{i}p_{i}=1$, and we indicate explicitly that we will consider entanglement properties

with respect to the $A|BC$ cut. Since we start from the state
$|\Psi^{+}\rangle_{AC_{1}}|\Psi^{+}\rangle_{C_{2}B}$, the Schmidt rank of each $\rho^i_{A|BC}$ cannot exceed $2$. Moreover, the entanglement of formation~\cite{Bennett3}
$E_F(\rho_{A|BCR})$ must be $1$, otherwise Alice and Bob (who is together with Charlie)
cannot obtain at the end a maximally entangled state, as required by ES. Using the fact that the states
$\rho^i_{A|BC}\otimes | i \rangle
\langle i |$ have support on locally orthogonal subspaces thanks to the classical registers, we obtain
$E_{F}=\sum_{i}p_{i}S(\rho_{A}^{i})$, with
$\rho_{A}^{i}={\rm Tr}_{BC}(\rho^i_{A|BC})$, and $S(\sigma)=-{\rm Tr}\sigma\log\sigma$ the von Neumann entropy.
We conclude that each state $\rho^i_{A|BC}$ is effectively
a maximally entangled state of two qubits, and $\rho_{A}^{i}={\openone}_2/2$ (on its local support), with ${\openone}_d$ the $d$-dimensional identity matrix. According to the previous remark, Alice's action on the state of the system is
\begin{equation}
\begin{split}
\Lambda_{A}(\rho_{ABCR})&=\sum_ip_i\Lambda_i(\rho^i_{A|BC})\otimes
|i \rangle \langle i |\\
&=\sum_{ij}p_{i}A_{ij}\rho_{A|BC}^{i}A_{ij}^\dag\otimes
|ij \rangle \langle ij |.
\end{split}
\end{equation}
Entanglement of formation becomes $E'_{F}=\sum_{ij}p_{i}q_{ij}S\Big(\frac{1}{q_{ij}}A_{ij}\rho_{A}^{i}A_{ij}^\dag\Big)$,
where ${q_{ij}}={\rm Tr}(A_{ij}\rho_{A}^{i}A_{ij}^\dag)$. Alice's operation $\Lambda_A$ can not be considered as a (probabilistic) isometry~\footnote{If the isometry is probabilistic, the choice of which isometry to apply (and the corresponding probability) may always be ascribed to Charlie's operation.} if at least one Kraus operator
$A_{i_{0}j_{0}}$ of hers does not act as an isometry (up to a coefficient of proportionality) on the local support. In such case $A_{i_{0}j_{0}}\rho_{A}^{i_{0}}A_{i_{0}j_{0}}^\dag$ is not proportional to ${\openone}_2$ and
$S(A_{i_{0}j_{0}}\rho_{A}^{i_{0}}A_{i_{0}j_{0}}^\dag/q_{i_{0}j_{0}})<
1$, therefore $E'_F<1$. Any further ($A|BC$)-bipartite LOCC processing of the state can not increase $E_F$, thus Alice and Bob cannot obtain at the end a maximally entangled state, and we reach a contradiction with the ES requirement. Therefore, neither Alice nor Bob can perform any other operation but isometries. Since their output subsystems correspond to their input subsystems, such isometries are indeed unitaries.

\emph{(II)} In all three cases Charlie is initially maximally entangled
with Alice and Bob with respect to his local support, i.e., two-qubit dimensional for ES and 2EPR-GHZ, and one-qubit dimensional for GHZ-EPR. Therefore, the action of the Kraus operator $E_C^i$ leads to the creation of a state whose Schmidt rank in the $(AB):C$ cut corresponds to the matrix rank of $E_C^i$. Thus, $E_C^i$ must have rank one, i.e., $E_C^i=\unidim{u^i}{\psi^i}$, with $\ket{\psi^i}$ a normalized state, for ES and GHZ-EPR boxes, and rank two, i.e., $E_C^i=c^i_0\unidim{u_0^i}{\psi_0^i}+c^i_1\unidim{u_1^i}{\psi_1^i}$, with $\braket{\psi_k^i}{\psi_l^i}=\braket{u_k^i}{u_l^i}=\delta_{kl}$, for all $i$,  for any 2EPR-GHZ box. 

\begin{description}
\item[{[ES]}] The output state of Alice's and Bob's subsystem corresponding to a given $E_C^i$ has the same Schmidt coefficients as $\ket{\psi_i}_{C_1C_2}$. Thus, we see that Alice and Bob may obtain a maximally
entangled state only if $\ket{\psi_i}$ is maximally entangled.

\item[{[2EPR-GHZ]}] it is immediate to check that $E_C^i\ket{\Psi^+}_{AC_1}\ket{\Psi^+}_{BC_2}=1/2(c_0\ket{u_0^i}_C\ket{\bar{\psi_0^i}}_{AB}+c_1\ket{u_1^i}_C\ket{\bar{\psi_1^i}}_{AB})$, with $\ket{\bar{\psi_k^i}}$ having the complex conjugate coefficients of $\ket{\psi_k^i}$ in the computational basis in which the EPR pairs are defined. In order for this state to be proportional to a GHZ state, it must hold $\ket{\psi_k^i}_{C_1C_2}=\ket{a_k^i}_{C_1}\ket{b_k^i}_{C_2}$ with $\braket{a_r^i}{a_s^i}=\braket{b_r^i}{b_s^i}=\delta_{rs}$, and $|c^i_0|=|c^i_1|$.

\item[{[GHZ-EPR]}] let us suppose that $\ket{\psi^i}=\cos\frac{\theta}{2}|0\rangle+e^{i\phi}\sin\frac{\theta}{2}|1\rangle$. Then the resulting state $\braket{\psi^i}{GHZ}$ of Alice's and Bob's subsystem, when normalized, has Schmidt coefficients $\cos\frac{\theta}{2}$ and $\sin\frac{\theta}{2}$. Thus, we see that Alice and Bob may obtain a maximally
entangled state only if $\ket{\psi_i}=\frac{1}{\sqrt{2}}(|0\rangle+e^{i\phi}|1\rangle)$

\end{description}

\end{proof}

\subsection{Upper and lower bounds on the communication cost}

We can now address the problem of the CC of ES-boxes. The standard form for ES-boxes of Theorem~\ref{thm:structure} tells us that all ES-boxes can be realized with only $C\rightarrow AB$ classical communication, which is used by Charlie to make Alice and Bob aware of the result of his measurement, so that they can apply the correct local unitary rotations.

In order to provide a lower bound on CC, we will need the entropic quantities $S(A|B)=S(AB)-S(B)$ (conditional entropy), $I(A:B)=S(A)+S(B)-S(AB)$ (mutual information), $I(A:B|R)=S(A|R)+S(B|R)-S(AB|R)$ (conditional mutual information), where for brevity we use the notation $S(X)=S(\rho_X)$, etc., and the following lemma.

\begin{lemma}
\label{lem:deltas}
Consider an ensemble $\{p^{i},\rho_{AB}^{i}\}$, and the corresponding average state $\rho_{AB}=\sum_ip^i\rho^i_{AB}$.
Then $\Delta I \leq \Delta S$, where $\Delta I= \sum_{i}p^{i} I
(\rho_{AB}^{i})-I(\rho_{AB})$ is the average increase of mutual
information, and $\Delta
S=S(\rho_{AB})-\sum_{i}p^{i}S(\rho_{AB}^{i})$ is the average decrease of
entropy, when Alice and Bob come to know the index of the state they actually share among the ones in the ensemble. 
\end{lemma}

\begin{proof}
Let us introduce a state $\rho_{ABR}=\sum_{i}
p^{i} \rho_{AB}^{i}\otimes|i \rangle_{R}\langle i|_{R}$, with orthogonal states $\ket{i}_R$, and calculate $I(A:B|R)$ for said state:
\beq
\begin{split}
I(A:B|R)& = S(AR)+S(BR)-S(R)-S(ABR)\\
&= S(\sum_{i}p^{i}\rho_{A}^{i}\otimes|i\rangle \langle i|)+S(\sum_{i}p^{i}\rho_{B}^{i}\otimes|i\rangle \langle i|)-S(\sum_{i}p^{i}|i\rangle \langle i|)-S(\sum_{i}p^{i}\rho_{AB}^{i}\otimes|i\rangle \langle i|)\\
& = H(\{p^{i}\})+ \sum_{i}p^{i}S(\rho_{A}^{i})+ H(\{p^{i}\})+ \sum_{i}p^{i}S(\rho_{B}^{i})-H(\{p^{i}\})- H(\{p^{i}\})- \sum_{i}p^{i}S(\rho_{AB}^{i})\\
& = \sum_{i}p^{i}(S(\rho_{A}^{i})+S(\rho_{B}^{i})-S(\rho_{AB}^{i}))\\
& = \sum_{i}p^{i}I(\rho_{AB}^{i}).
\end{split}
\eeq
Thus, we have obtained the expression
\begin{eqnarray}
\Delta I = I(A:B|R)-I(A:B).
\end{eqnarray}
for the average increase of mutual information.
Let us further calculate $I(AB:R)$ for $\rho_{ABR}$:
\beq
\begin{split}
I(AB:R)&=S(\rho_{AB})+S(\rho_{R})-S(\rho_{ABR})\\
& =S(\rho_{AB})+H(\{p^{i}\})-H(\{p^{i}\})-\sum_{i}p^{i}S(\rho_{AB})\\
& =S(\rho_{AB})-\sum_{i}p^{i}S(\rho^i_{AB}).
\end{split}
\eeq
Therefore, the average decrease of entropy is given by
\begin{eqnarray}
\Delta S = I(AB:R).
\end{eqnarray}
We are now ready to show that $\Delta I \leq \Delta S$:
\begin{equation}
\begin{split}
\Delta I - \Delta S &= I(A:B|R)-I(A:B)-I(AB:R)\\
&=S(AR)+S(BR)-S(R)-S(ABR)-S(A)-S(B)+S(AB)-S(AB)-S(R)+S(ABR)\\
&=-[S(A)+S(R)-S(AR)]-[S(A)+S(R)-S(AR)]\\
&= -I(A:R)-I(B:R)\\
&\leq 0
\end{split}
\end{equation}
The last inequality holds because of non-negativity of mutual information.
\end{proof}

Since Charlie has to tell Alice and Bob the result of his measurement, the number of bits sent by Charlie to Alice and Bob per each realization of a ES-box can not be less than the Shannon entropy $H(\{p_i\})=-\sum_ip_i\log p_i$ of the probability distribution of the outcomes of Charlie. This line of reasoning leads to the following CC bounds. 

\begin{theorem}
The following bounds hold for any box realizing the corresponding task:
\begin{description}
\item[{[ES]}] $\rm{CC}\geq 2$;
\item[{[2EPR-GHZ]}] $\rm{CC}\geq 1$;
\item[{[GHZ-EPR]}] $\rm{CC}\geq 1$.
\end{description}
\end{theorem}

\begin{proof}
Starting with the right input, before Charlie's measurement, the reduced state $\rho_{AB}$ is the maximally mixed state ${\openone}_4/4$ for both an ES-box and a 2EPR-GHZ box, while it is a maximally classically correlated state $\frac{1}{2}(\pro{00}+\pro{11})$ for a GHZ-EPR box. The mutual information is $I(\rho_{AB})=0$ in the first two cases and $I(\rho_{AB})=1$ in the last case.
After Charlie's measurement, Alice and Bob have
an ensemble $\{p^{i},\rho_{AB}^{i}\}$. Each $\rho_{AB}^i$ is a maximally entangled state for ES-box or GHZ-EPR box and a maximally classically correlated state for 2EPR-GHZ box. The mutual information is $I(\rho_{AB}^{i})=2$ in the first two cases and $I(\rho_{AB}^{i})=1$ in the last case. From Lemma~\ref{lem:deltas} we have
\beq
\begin{split}
H(\{p_{i}\})&=S(\rho_{AB})-\sum_{i}p^{i}S(\rho_{AB}^{i})\\
&\geq\sum_{i}p^{i}I(\rho_{AB}^{i})-I(\rho_{AB}).
\end{split}
\eeq
Hence, we obtain the lower bounds $H(\{p_{i}\})=2$ for ES box, and $H(\{p_{i}\})=1$ for both 2EPR-GHZ box and GHZ-EPR box.
The protocols which achieve these bounds are:
\begin{description}
\item[{[ES]}] standard teleportation map: Bell measurement of Charlie, and conditional Pauli rotation on Alice and/or Bob;
\item[{[2EPR-GHZ]}] Charlie's measurement $\{ |00\rangle \langle 00|+|11\rangle \langle 11|, |01\rangle \langle 01|+|10\rangle \langle 10| \}$, and conditional unitary operation on Bob's qubit $\{\openone, X\}$;
\item[{[GHZ-EPR]}] Charlie's measurement $\{ \frac{1}{2}(|0\rangle+ |1\rangle) (\langle 0|+\langle 1|),  \frac{1}{2}(|0\rangle- |1\rangle) (\langle 0|-\langle 1|)\}$, and conditional unitary operation on Bob's qubit $\{\openone, Z\}$.
\end{description}
\end{proof}

Thus, we have obtained lower bounds for the communication cost of any entanglement redistribution boxes. Furthermore, we proved that such bounds are tight, in the sense that there exist boxes that require exactly that amount of $C\rightarrow AB$ communication to be realized. Thus, $CC=1$ for both 2EPR-GHZ and GHZ-EPR, and $CC=2$ for ES.

\subsection{Upper and lower bounds on the communication value}

We now look for lower bounds on how useful to establish communication among the users a box is. In order to do this, we will provide schemes to exploit the entanglement redistribution boxes to communicate. Let us remark that one may consider two different scenarios. In the first one, the box is really black to the users, i.e., they do not know which box, among all the possible ones realizing a particular task, they are dealing with. In the second one, the users actually know which particular box realizing a certain task they have been provided with, so they can better exploit it for communication. In the latter case, the box might not be considered to be ``black'', as users know its content. Anyway, on the one hand, having bounds which are valid for any box realizing a certain task, implies that each black box comes with a certain potential for communication, which can be ``activated'' if at a certain point the users come to know which box they are dealing with -- e.g., through process tomography. On the other hand, as it will turn out, the protocols we devise to communicate do depend only on the defining process of entanglement redistribution, and on the structural characterization of Theorem~\ref{thm:structure}. Thus, the same protocol works for any box realizing a precise task, even if it is black.

The following theorem provides a lower bound for CV for ES and GHZ-EPR boxes.
 
\begin{theorem}
\label{thm:lowerboundCV}
Any ES-box or GHZ-EPR box has CV at least equal to 1.
\end{theorem}

\begin{proof}
We use two EPR pairs $|\Psi^{+}\rangle$ as an input to the box. In order to communicate to Alice and Bob Charlie applies one of two possible unitary operations, $I$ or $Z_{C1}$, to his part of the input, before the action of the box. We can write the result of the action of a ES-box on two EPR pairs in the following way
\beq
\begin{split}
\label{eq:ESout}
\Psi^{+}_{AB}&=Tr_C\Lambda_{ABC}(\Psi^{+}_{AC_{1}}\Psi^{+}_{BC_{2}})\\ 
\Lambda_{ABC}(\Psi^{+}_{AC_{1}}\Psi^{+}_{BC_{2}})&=\sum_{i}U_{A}^{i}U_{B}^{i}E_{C}^{i}\Psi^{+}_{AC_{1}}\Psi^{+}_{BC_{2}} U_{A}^{i\dag}U_{B}^{i\dag}E_{C}^{i\dag}
\end{split}
\eeq
Similarly, the the action of an GHZ-EPR box on an input given by the GHZ state is
\beq
\label{eq:GHZ-EPRout}
\begin{split}
\Psi^{+}_{AB}&=Tr_{C}\Lambda_{ABC}(|GHZ \rangle\langle GHZ|_{ABC})\\
\Lambda_{ABC}(|GHZ \rangle\langle GHZ|_{ABC}) & =\sum_{i}U_{A}^{i}U_{B}^{i}E_{C}^{i}
(|GHZ \rangle\langle GHZ|_{ABC})U_{A}^{i\dag}U_{B}^{i\dag}E_{C}^{i\dag}
\end{split}
\eeq

Let us suppose that, before the action of $\Lambda_{ABC}$, Charlie applies, in the ES case, $Z_{C_1}$ to his qubit of the first EPR pair for the ES-box, so that the $AB$ output is
\begin{equation}
\rho_{AB}=Tr_{C}\sum_{i}U_{A}^{i}U_{B}^{i}E_{C}^{i}Z_{C_{1}}
\Psi^{+}_{AC_{1}}\Psi^{+}_{BC_{2}}Z_{C_{1}}^{\dag}E_{C}^{i\dag}U_{A}^{i\dag}U_{B}^{i\dag},
\end{equation}
Similarly, in the GHZ-EPR box case, if Charlie applies $Z_{C}$ to his qubit of the GHZ state, the $AB$ output is
\begin{equation}
\rho_{AB}=Tr_{C}\sum_{i}U_{A}^{i}U_{B}^{i}E_{C}^{i}Z_{C}
|GHZ\rangle\langle GHZ|_{ABC}Z_{C}^{\dag}E_{C}^{i\dag}U_{A}^{i\dag}U_{B}^{i\dag}
\end{equation}
We will now show that the output $\rho_{AB}$ of the box is orthogonal to $|\Psi^{+}\rangle_{AB}$ in both cases.

Let us first note the two identities
\begin{equation}
Z_{C_{1}}|\Psi^{+}\rangle_{AC_{1}}=Z_{A}|\Psi^{+}\rangle_{AC_{1}}\quad Z_{C}|GHZ\rangle = Z_{A}|GHZ\rangle.
\end{equation}
Using said identities, we can rewrite the output $\rho_{AB}$ as
\begin{equation}
\label{eq:output}
\rho_{AB}=\sum_{i}U_{A}^{i}U_{B}^{i}Z_{A}\rho_{AB}^{i}
Z_{A}^{\dag}U_{A}^{i\dag}U_{B}^{i\dag}
\end{equation}
with
\begin{equation}
\rho_{AB}^{i}=Tr_{C}E_{C}^{i}
\Psi^{+}_{AC_{1}}\Psi^{+}_{BC_{2}}E_{C}^{i\dag}
\end{equation}
for ES-box and
\begin{equation}
\rho_{AB}^{i}=Tr_{C}E_{C}^{i}
|GHZ\rangle \langle GHZ|_{ABC}E_{C}^{i\dag}
\end{equation}
for GHZ-EPR box.

From definition of ES-box and GHZ-EPR box, i.e., Eqs.~\eqref{eq:ESout} and~\eqref{eq:GHZ-EPRout}, in both cases we have
\begin{equation}
\Psi^{+}_{AB}=\sum_{i}U_{A}^{i}U_{B}^{i}\rho_{AB}^{i}U_{A}^{i\dag}U_{B}^{i\dag},
\end{equation}
which means that for all $i$ corresponding to an output with non-vanishing probability, we have
\begin{equation}
\label{eq:propmaxent}
U_{A}^{i}U_{B}^{i}\rho_{AB}^{i}U_{A}^{i\dag}U_{B}^{i\dag}=r_{i}\Psi^{+}_{AB}
\end{equation}
for some strictly positive parameter $r_i$.

Let us now calculate in both cases the overlap between the output state $\rho_{AB}$ and $\Psi^{+}_{AB}$. We have
\beq
\begin{split}
Tr(\rho_{AB}\Psi^{+}_{AB})&=\sum_{i}Tr(U_{A}^{i}U_{B}^{i}Z_{A}\rho_{AB}^{i}Z_{A}^{\dag}U_{A}^{i\dag}U_{B}^{i\dag}\Psi^{+}_{AB})\\
& =\sum_{i}Tr(Z_{A}\rho_{AB}^{i}Z_{A}^{\dag}U_{A}^{i\dag}U_{B}^{i\dag}\Psi^{+}_{AB}U_{A}^{i}U_{B}^{i})\\
& =\sum_{i}\frac{1}{r_{i}}Tr(Z_{A}\rho_{AB}^{i}Z_{A}^{\dag}\rho_{AB}^{i})
\end{split}
\eeq
where we have used: Eq. \eqref{eq:output} in the first equality; the cyclic property of trace in the second equality; Eq. \eqref{eq:propmaxent} in the third equality.

We will now prove that $Tr(Z_{A}\rho_{AB}^{i}Z_{A}^{\dag}\rho_{AB}^{i})$ vanishes for all $i$. Since $\rho_{AB}^{i}$ is proportional to a maximally entangled state (Eq. \eqref{eq:propmaxent}) we can write 
\begin{equation}
\rho_{AB}^{i}=r_{i}V_{B}^{i}\Psi^{+}_{AB}V_{B}^{i \dag},
\end{equation}
where $V_{B}^{i}$ is some unitary.
We thus have
\beq
\begin{split}
Tr(Z_{A}\rho_{AB}^{i}Z_{A}^{\dag}\rho_{AB}^{i})&=r_{i}^{2}|\langle\Psi^{+}|_{AB}Z_{A}|\Psi^{+}\rangle_{AB}|^{2}\\
																								&=\frac{r_{i}^{2}}{4}\tr(Z_{A})^2\\
																								&=0
\end{split}
\eeq
\end{proof}

It should again be emphasized that, in order to communicate, Charlie can apply operations that do not depend on the particular ES or GHZ-EPR box, so that communication is achieved whatever the black box at disposal, i.e. the internal structure of the box -- the particular LOCC map -- is not relevant.

We will now prove that the bound $\rm{CV}=1$ can be achieved, i.e., there exist boxes such that the user Charlie can not send to users Alice and Bob more than one bit per use of the box. The maps we provide to this purpose happen to be also $C \rightarrow A$ and $C \rightarrow B$ nonsignaling, i.e. Charlie cannot communicate to neither Alice nor Bob separately. We will use the standard bipartite operation of $UU^*$-twirling $\Lambda_{AB}^{T}(\sigma_{AB})\equiv\int dU(U_A\otimes U^{\ast}_B)\sigma_{AB} (U_A \otimes U^{\ast}_B)^{\dag}$, where $\int dU$ denotes integration over the unitary group with respect to the Haar measure, and $U^\ast$ is the complex conjugate of $U$. 

\begin{theorem}
\label{thm:CV}
Apply the $UU^*$-twirling to the output of any ES (GHZ-EPR) box $\Lambda_{ABC}$: the resulting map is again an ES (GHZ-EPR) box , with CV equal to $1$, and non-signaling with respect to $C \rightarrow A$ and $C \rightarrow B$.

\end{theorem}

\begin{proof} 

The action $UU^*$-twirling map is given by 
\beq
\begin{split}
\Lambda_{AB}^{T}(\rho_{AB})&=\int dU U\otimes U^{\ast}\sigma U^{\dag} \otimes U^{T}\\
				&=F\Psi^{+}_{AB} + \frac{1-F}{3}(\openone_{AB}-\Psi^{+}_{AB}),
\end{split}
\eeq
where $F=\langle\Psi^{+}|\rho_{AB}|\Psi^{+}\rangle_{AB}$. It is clear that this map leaves $\Psi^{+}_{AB}$ invariant and hence the map $\tilde{\Lambda}_{ABC}=\Lambda_{AB}^{T}\circ\Lambda_{ABC}$ is still an ES (GHZ-EPR) box.

As the parties may take advantage of pre-shared entanglement in order to communicate via the box, we will use the Entanglement-Assisted Classical Capacity of a Quantum Channel (EACCQC)~\cite{Bennett7} to bound from above communication value
of the channel $\tilde{\Lambda}_{ABC}$. Notice that the standard notion of EACCQC~\cite{Bennett7} applies to a channel with one-party input and one-party output, i.e. the input of the channel is completely controlled by the sender. Although we have reduced our attention to the $C\rightarrow AB$ communication, i.e., one receiver and one sender, still the channel is $C:AB$ bipartite, so that the setting we are studying does not fit properly in the standard framework.   Anyway, consistently with the idea of obtaining an upper bound, we apply the formula for EACCQC by pretending that Charlie has complete control over all the input, not only on his subsystem, so that it reads
\beq
\label{eq:EACCQC}
C(\tilde{\Lambda}_{ABC})= \text{max} (S(\rho_{ABC}) + S(\tilde{\Lambda}_{ABC}(\rho_{ABC})) -S((\tilde{\Lambda}_{ABC}\otimes \id_{E})(\Phi_{ABCE}))),
\eeq
where $\Phi_{ABCE}$ is any purification of $\rho_{ABC}$, $\id$ is the identity map, and the maximum
runs  over all possible input states $\rho_{ABC}$. 
For our channel we have
\beq
\tilde{\Lambda}_{ABC}(\rho_{ABC})=F \Psi^{+}_{AB} +\frac{1-F}{3}(\openone_{AB}-\Psi^{+}_{AB})
\eeq
and 
\beq
(\tilde{\Lambda}_{ABC}\otimes \id_{E})(\Phi_{ABCE})= F\Psi^{+}_{AB} \otimes \rho^{0}_{E} + \frac{1-F}{3}(\openone_{AB}-\Psi^{+}_{AB}) \otimes \rho^{1}_{E}
\eeq
where $F=\langle \Psi^{+}_{AB}|\Lambda_{ABC}(\rho_{ABC})|\Psi^{+}\rangle_{AB}$, and $\rho_{E_0}$ and $\rho_{E_1}$ are two states of the environment.

We will now calculate the values of the various entropies entering the formula for EACCQC. First, we have
\beq
\begin{split}
S(\rho_{ABC})&=S(\rho_{E})\\
&= S(Tr_{ABC}(\tilde{\Lambda}_{ABC}\otimes \id_{E}(\Phi_{ABCE})))\\
&= S(F\rho_{E_{0}}+(1-F)\rho_{E_{1}})
\end{split}
\eeq
The first equality comes from the fact that the state of the whole system -- reference system $E$ included -- is pure, and the second equality comes from the fact that the box does not affect the state of the environment.

Second, we find
\beq
S(\tilde{\Lambda}_{ABC}(\rho_{ABC}))=H(F)+(1-F)\log3,
\eeq
where $H(x)\equiv -x \log x - (1-x) \log (1-x)$ is the binary entropy.

Third,  we compute
\beq
\begin{split}
S((\tilde{\Lambda}_{ABC}\otimes \id_{E})(\Phi_{ABCE})))
& =S\Big(F\Psi^{+}_{AB}\otimes \rho^{0}_{E} + \frac{1-F}{3}(\openone_{AB}-\Psi^{+}_{AB}) \otimes \rho^{1}_{E}\Big)\\
& = H(F)+ FS(\Psi^{+}_{AB} \otimes \rho^{0}_{E})+(1-F)S\big(\frac{1}{3}(\openone_{AB}-\Psi^{+}) \otimes \rho^{1}_{E}\big)\\
& =H(F)+FS(\rho_E^{0})+(1-F)S(\rho_E^{1})+(1-F)\log3
\end{split}
\eeq
The second equality comes from the fact that states $\Psi^{+}_{AB} \otimes \rho^{0}_{E}$ and $\frac{1}{3}(\openone_{AB}-\Psi^{+}_{AB}) \otimes \rho^{1}_{E}$ have support on orthogonal subspaces.

Finally, by substituting the above entropies into the formula for EACCQC we obtain
\beq
C(\tilde{\Lambda}_{ABC})= \max
\Big( S\big(F\rho^{0}_{E}+(1-F)\rho^{1}_{E}\big) - \big(F S(\rho^{0}_{E})+(1-F) S(\rho^{1}_{E}))\big) \Big) \leq H(F) \leq 1
\eeq
Thus we bounded from the above communication value of the map. On
the other hand from Theorem \ref{thm:lowerboundCV} we know that any ES (GHZ-EPR) box has
communication value at least 1 with respect to $C \rightarrow AB$.
Moreover $\tr_{B(A)}(\tilde{\Lambda}_{ABC}(\rho_{ABC}))=\frac{1}{2}\openone_{A(B)}$ does
not depend on Charlie's action $\Gamma_{C}^{i}$, which means that the
map is $C \rightarrow A(B)$ {\it nonsignaling}.
\end{proof}

%\textbf{UNTIL HERE + PART OF CONCLUSIONS}

We also note that the map $\tilde\Lambda_{ABC}$, as defined in Theorem \ref{thm:CV}, has CC less or equal to that of the map $\Lambda_{ABC}$ from which it derives. This follows from the fact that twirling can be performed without communication, by means of shared randomness.
%We may take $\Lambda_{ABC}$ to be standard teleportation, so that $\tilde\Lambda_{ABC}$ has CC equal to 2 and CV equal to 1.
%Thus both CC and CV of this specific ES-box coincide with those of ES.

Thus, we have obtained that ES and GHZ-EPR have, as tasks, a CV equal to one. In practical terms, this means that when the users are given any ES or GHZ-EPR box, they know how -- the protocol to communicate is independent of the box -- and how much it is \emph{for sure} possible to communicate from Charlie to Alice and Bob.

As regards the redistribution 2EPR-GHZ, we are only able to provide a weaker result: it can be used to communicate more than $\approx 0.3219$ bits per use.

\begin{theorem}
Any 2EPR-GHZ box has CV strictly greater than $0$, in particular it holds $CV\gtrsim 0.3219$.
\end{theorem}
\begin{proof}
We use two EPR pairs $|\Psi^{+}\rangle$ as an input to the box. In order to communicate to Alice and Bob Charlie applies one of two operations: either the identity or a totally depolarizing random unitary, $D(X_C)\equiv\mathcal{N}\sum_{i} U_{C}^{i}X_{C}U_{C}^{i\dagger}=tr(X)\openone_{C}/4$, for all $X_{C}$with $\mathcal{N}$ a normalization to make the map $D$ trace preserving.
We show that the corresponding Alice and Bob's reduced density operators of the output of the box are different -- although non-orthogonal -- quantum states, hence they can be used to communicate.  We can write the action of 2EPR-GHZ on two EPR pairs in the following way
\begin{equation}
\label{eq:2EPR-GHZ}
\begin{split}
|GHZ\rangle\langle GHZ|_{ABC}&=\Lambda_{ABC}(\Psi^{+}_{AC_{1}}\Psi^{+}_{BC_{2}})\\
								& =\sum_{i}U_{A}^{i}U_{B}^{i}E_{C}^{i} \Psi^{+}_{AC_{1}}\Psi^{+}_{BC_{2}}E_{C}^{i\dag}U_{A}^{i\dag}U_{B}^{i\dag}
\end{split}
\end{equation}
and Alice and Bob's reduced density matrix is the maximally classically correlated state
\begin{eqnarray}
\rho_{AB}=\frac{1}{2}(|0\rangle_{A}|0\rangle_{B}\langle0|_{A}\langle0|_{B}+|1\rangle_{A}|1\rangle_{B}\langle1|_{A}\langle1|_{B})
\end{eqnarray}
Let us suppose that before $\Lambda_{ABC}$ Charlie performs the totally depolarizing random unitary $D$ on his qubits, i.e., the output state of Alice and Bob is in this case
\begin{equation}
\overline{\rho}_{AB}=Tr_{C}\sum_{i}\sum_{j}U_{A}^{i}U_{B}^{i}E_{C}^{i}U_{C}^{j}
\Psi^{+}_{AC_{1}}\Psi^{+}_{BC_{2}}U_{C}^{j\dagger}E_{C}^{i\dag}U_{A}^{i\dag}U_{B}^{i\dag}
\end{equation}
We will show that $\overline{\rho}_{AB}$ is different from $\rho_{AB}$. We have
\begin{eqnarray}
\sum_{j}U_{C}^{j}
\Psi^{+}_{AC_{1}}\Psi^{+}_{BC_{2}}U_{C}^{j\dag}=\frac{1}{4}\openone_{AB}\frac{1}{4}\openone_{C}
\end{eqnarray}
and hence
\[
\begin{split}
\overline{\rho}_{AB}&=\frac{1}{16}Tr_{C}\sum_{i}U_{A}^{i}U_{B}^{i}E_{C}^{i}E_{C}^{i\dag}U_{A}^{i\dag}U_{B}^{i\dag}\\
		&=\frac{1}{16}Tr_{C}(\sum_{i}E_{C}^{i\dag}E_{C}^{i})\openone_{AB}\\
		&=\frac{1}{4}\openone_{AB}
\end{split}
\]
In the second equality we used the cyclic property of trace and in the third equality we used that $\sum_{i}E_{C}^{i\dag}E_{C}^{i}=\openone_{C}$.

Let us now calculate the Holevo quantity for the ensemble $\{(p,\rho_{AB}),(1-p,\overline{\rho}_{AB})\}$. We have
\[
\begin{split}
\chi&=S(p\rho_{AB}+(1-p)\overline{\rho}_{AB})-pS(\rho_{AB})-(1-p)S(\overline{\rho}_{AB})\\
	&=S\Big(\frac{1}{4}(1+p)(|00\rangle\langle00|_{AB}+|11\rangle\langle11|_{AB})+\frac{1}{4}(1-p)(|01\rangle\langle01|_{AB}+|10\rangle\langle10|_{AB})\Big)\\
	&-pS\Big(\frac{1}{2}|00\rangle\langle00|_{AB}+\frac{1}{2}|11\rangle\langle11|_{AB}\Big)-(1-p)S\Big(\frac{1}{4}\openone_{AB}\Big)\\
& =H(\frac{1+p}{2})+p-1
\end{split}
\]
Maximizing over $p$ we obtain $\chi_{\text{max}}\approx 0.3219$ for $p=0.6$. Hence using this protocol Charlie can communicate to Alice and Bob $0.3219$ bits.
\end{proof}

The previous theorem provides a lower bound for the CV $C\rightarrow AB$ of any 2EPR-GHZ box, therefore also for the CV of the 2EPR-GHZ task. As done for the other two entanglement redistribution tasks, we now provide an upper bound on the CV, which in this case does not coincide with the lower bound. Thus are only able to provide an interval for the CV of the 2EPR-GHZ task. 

\begin{theorem}
Apply the $U_{A}V_{B}U_{C_1}^*V_{C_2}^*$-twirling $\Lambda_{AC_1}^{T} \otimes  \Lambda_{BC_1}^{T}$ to the input of some 2EPR-GHZ box $\Lambda_{ABC}$: the resulting map $\tilde{\Lambda}_{ABC}={\Lambda}_{ABC}\circ (\Lambda_{AC_1}^{T} \otimes  \Lambda_{BC_1}^{T})$ is again an 2EPR-GHZ box, with CV less than $1$, and non-signaling with respect to $C \rightarrow A$ and $C \rightarrow B$. In particular, it holds $CV\lesssim 0.469782$.
\end{theorem}

\begin{proof}
The action $U_{A}V_{B}U_{C_1}^*V_{C_2}^*$-twirling map on the input state is given by 
\beq
\label{eq:doubletwirl}
\begin{split}
(\Lambda_{AC_1}^{T} \otimes  \Lambda_{BC_1}^{T})(\rho_{ABC})
&=
A \Psi^{+}_{AC_1}\otimes \Psi^{+}_{BC_2} \\ 
& + B\Psi^{+}_{AC_1} \otimes \frac{1}{3}(\openone_{BC_2}-\Psi^{+}_{BC_2})\\ 
& +C\frac{1}{3}(\openone_{AC_1}-\Psi^{+}_{AC_1})\otimes \Psi^{+}_{BC_2}\\ 
& +D\frac{1}{3}(\openone_{AC_1}-\Psi^{+}_{AC_1})\otimes\frac{1}{3}(\openone_{BC_2}-\Psi^{+}_{BC_2})
\end{split}
\eeq
where
\beq
\label{eq:doubletwirl_par}
\begin{aligned}
A &= Tr (\Psi^{+}_{AC_1}\otimes \Psi^{+}_{BC_2} \rho_{ABC})\\ 
B &= Tr (\Psi^{+}_{AC_1} \otimes (\openone_{BC_2}-\Psi^{+}_{BC_2})\rho_{ABC})\\ 
C &= Tr ((\openone_{AC_1}-\Psi^{+}_{AC_1})\otimes \Psi^{+}_{BC_2}\rho_{ABC})\\ 
D &= Tr ((\openone_{AC_1}-\Psi^{+}_{AC_1})\otimes(\openone_{BC_2}-\Psi^{+}_{BC_2})\rho_{ABC}),
\end{aligned}
\eeq
with $A+B+C+D=1$, because of normalization.

We observe that, by the defining action \eqref{eq:2EPR-GHZ} of a 2EPR-GHZ box, we must have
\[
U_{A}^{i}U_{B}^{i}E_{C}^{i} \ket{\Psi^{+}}_{AC_{1}}\ket{\Psi^{+}}_{BC_{2}}=s_i \ket{GHZ}_{ABC}\quad \forall i
\]

with $s_i$ some proportionality constant such that $\sum_i |s_i|^2=1$.
Suppose that the input $\ket{\Psi^{+}}_{AC_{1}}\ket{\Psi^{+}}_{BC_{2}}$ is substituted by $\sigma^A_\mu \sigma^B_\nu\ket{\Psi^{+}}_{BC_{2}}\ket{\Psi^{+}}_{BC_{2}}$, with $\sigma_\mu, \sigma_\nu$ Pauli matrices. Then, it holds
\[
\begin{split}
(U_{A}^{i}U_{B}^{i}E_{C}^{i}) (\sigma^A_\mu \sigma^B_\nu)\ket{\Psi^{+}}_{BC_{2}}\ket{\Psi^{+}}_{BC_{2}}
&=(U_{A}^{i}\sigma^A_\mu U_{A}^{i\dag})(U_{B}^{i}\sigma^B_\nu U_{B}^{i\dag})U_{A}^{i}U_{B}^{i}E_{C}^{i} \ket{\Psi^{+}}_{AC_{1}}\ket{\Psi^{+}}_{BC_{2}}\\
&=s_i (U_{A}^{i}\sigma^A_\mu U_{A}^{i\dag})(U_{B}^{i}\sigma^B_\nu U_{B}^{i\dag})\ket{GHZ}_{ABC},
\end{split}
\]
i.e. for each branch of the protocol, the output is a locally rotated GHZ state. Let us further observe that the maximally mixed state of two qubits can be written as the convex combination of four maximally entangled states, i.e.,
\[
\openone_{BC_2}=\sum_{\nu=0}^3 \Psi^\mu_{BC_2},\quad \Psi^\mu_{BC_2}=\pro{\Psi^\mu}_{BC_2}, \quad\ket{\Psi^\mu}_{BC_2}=\sigma^B_\mu \ket{\Psi^{+}}_{BC_{2}}.
\]
Suppose now that the input of the original 2EPR-GHZ box is $\Psi^{+}_{AC_1}\otimes \frac{\openone_{BC_2}}{4}$. Putting together the previous observations, we have
\[
\begin{split}
\Lambda_{ABC}\Big[\Psi^{+}_{AC_1}\otimes \frac{\openone_{BC_2}}{4}\Big]&= \frac{1}{4}\sum_i\sum_\nu |s_i|^2
	  (U_{B}^{i}\sigma^B_\nu U_{B}^{i\dag})\ket{GHZ}\bra{GHZ}_{ABC}(U_{B}^{i}\sigma^B_\nu U_{B}^{i\dag})\\
	&= \sum_i |s_i|^2 \frac{1}{4}\sum_\nu (U_{B}^{i}\sigma^B_\nu U_{B}^{i\dag})\ket{GHZ}\bra{GHZ}_{ABC}(U_{B}^{i}\sigma^B_\nu U_{B}^{i\dag})\\
	&= \sum_i |s_i|^2  U_{B}^{i} \Big[Tr_B\big(U_{B}^{i\dag}\ket{GHZ}\bra{GHZ}_{ABC} U_{B}^{i}\big)\otimes \frac{\openone_B}{2}\Big] U_{B}^{i\dag}\\
	&= \sum_i |s_i|^2  \frac{1}{2}(\pro{00}+\pro{11})_{AC}\otimes \frac{\openone_B}{2}\\
	&= \frac{1}{2}(\pro{00}+\pro{11})_{AC}\otimes \frac{\openone_B}{2},
\end{split}
\]
where in the third equality we used that
\[
\frac{1}{4}\sum_\nu \sigma_\nu X \sigma_\nu = Tr(X) \frac{{\openone}_2}{2} \quad \forall X.
\]
Hence
\beq
\label{eq:halfmixed}
 Tr_C\Big(\Lambda_{ABC}\Big[\Psi^{+}_{AC_1}\otimes \frac{\openone_{BC_2}}{4}\Big]\Big)=\frac{\openone_{AB}}{4}.
\eeq
Similarly, one checks that
\beq
\label{eq:allmixed}
 Tr_C\Big(\Lambda_{ABC}\Big[\frac{\openone_{AC_1}}{4}\otimes\Psi^{+}_{BC_2}\Big]\Big)= Tr_C\Big(\Lambda_{ABC}\Big[\frac{\openone_{AC_1}}{4}\otimes\frac{\openone_{BC_2}}{4}\Big]\Big)=\frac{\openone_{AB}}{4}. 
\eeq
By combining the Eqs. \eqref{eq:doubletwirl}, \eqref{eq:halfmixed}, and \eqref{eq:allmixed}, with some further algebra we arrive at
\beq
\label{eq:ABoutput}
Tr_C((\tilde{\Lambda}_{ABC})(\rho_{ABC}))=F'\rho_{even}+(1-F')\rho_{odd},
\eeq
where
\beq
\label{eq:Fpar}
F'=A+\frac{B+C}{3}+\frac{5D}{9}
\eeq
and
\be
\rho_{even} = \frac12 (\pro{00}+\pro{11}),\quad \rho_{odd} = \frac12 (\pro{10}+\pro{01}).
\ee
Recall that the only constraint for $A,B,C,D$ is 
that they sum up to 1. 

Now we will use Lemma~\ref{lem:EACCequalsCC} (proven below), which tells us that, 
if a channel is of the form 
$\Lambda=\Gamma' \circ \Gamma$, where $\Gamma$ projects onto abelian algebra, then its classical capacity 
is equal to its entanglement assisted classical capacity. 
In our case $\Gamma$ is (double) twirling, hence it projects onto abelian algebra spanned by projectors 
$\Psi_+\ot \Psi_+,(\openone-\Psi_+)\ot \Psi_+,\Psi_+\ot (\openone-\Psi_+),(\openone-\Psi_+)\ot (\openone-\Psi_+)$.
It follows that it is enough to compute Holevo capacity 
of the channel (which is additive in this case).

We shall again pretend, that all the initial input of $\tilde{\Lambda}_{ABC}$ is in the hands of Charlie. Then, according to Eqs. \eqref{eq:ABoutput} and \eqref{eq:Fpar}, the possible AB output states are all the states of the form
\[
\rho_{F'}=F'\rho_{even}+(1-F')\rho_{odd},
\]
with $F'$ in the range $1/3\leq F'\leq 1$, as $A,B, C,D\geq0$ and $A+B+C+D=1$. Therefore, the maximal Holevo quantity is
\[
\chi_{\max}=\max_{p_i,F'_i}\Big(S(\sum_i p_i \rho_{F'_i}) - \sum_i p_i S(\rho_{F'_i})\Big),
\]
which, considering that $\rho_{even}$ and $\rho_{odd}$ are orthogonal and that $S(\rho_{even})=S(\rho_{odd})=1$, can be expressed as
\[
\chi_{\max}=\max_{p_i,F'_i}\Big(H(\sum_i p_i F'_i) - \sum_i p_i H(F'_i)\Big).
\]
In turn, by the concavity of entropy, one finds that only the extreme values $F'=1/3$ and $F'=1$ must be considered, so that
\be
\chi_{\max}=\sup_{p}\chi(\{(p,\rho_{1}),(1-p,\rho_{1/3}) \} )\simeq 0.469782,
\ee
where $\chi\bigl(\{(p,\rho_1),(1-p,\rho_{1/3})\}\bigr)=H({2p+1\over 3})-(1-p)H(\frac13)$.
\end{proof}

\begin{lemma}
\label{lem:EACCequalsCC}
For a channel of the form $\Lambda=\Gamma' \circ \Gamma$,  where $\Gamma$ is 
a projection onto an abelian algebra, the classical capacity  and the entanglement assisted classical capacity are equal.
\end{lemma}
\begin{proof}
For a channel $\Lambda$ acting on system $A$ 
and producing output on system $B$, the formula for EACCQC is given by 
\be
C_E(\Lambda)=\sup_{\rho_{EA}}I\bigl((\id_E\ot \Lambda_A)(\rho_{EA}) \bigr),
\ee
where the supremum runs over all states $\rho_{EA}$,
with $E$ being of the same dimension as $A$
(it is actually enough to consider pure states). 
The formula for Holevo capacity can be written as 
\be
C_1(\Lambda)=\sup_{\sigma_{EA}}I\bigl((\id_E\ot \Lambda_A)(\sigma_{EA}) \bigr)
\label{eq:clas-cap}
\ee
where the supremum runs over all $\sigma_{EA}$ of the c-q form 
\be
\sigma_{EA}=\sum_i p_i |i\>_E\<i| \ot \sigma^i_A
\ee
for any size of system $E$ (via Caratheodory theorem 
it is known that a finite one is enough, but 
it is not important for the moment). 
Since clearly $C_1\leq C_E$, it is enough to show the converse inequality. To this end, we shall show that 
for any state $\rho_{EA}$ one can build a c-q state $\sigma_{EA}$ 
such that after action of channel, 
the latter state has no smaller mutual information 
than the former one. Indeed, fix any  state $\rho_{EA}$.
Then, for our channel $\Lambda$ we have 
\be
(\id_E\ot \Lambda_A)(\rho_{EA})=(\id_E\ot \Gamma'_A)\bigl(\sum_i p_i \rho_E^i\ot P^i_A\bigr) 
\ee
where $P^i$'s are orthogonal projectors from the abelian algebra. Now, notice 
that mutual information can only increase, if we take the following state 
\be
\sum_i p_i |i\>_{E'}\<i|\ot P^i_A,
\label{eq:cq-state2}
\ee
because the state $\sum_i p_i \rho_E^i\ot P^i_A$ can be obtained 
from it by action on system $E$ solely. Since this state is preserved by $\Gamma$ 
we get that  
\be
C_E(\Lambda)=\sup (\id_E\ot \Lambda_A)(\rho_{EA}),
\ee
where the supremum is taken over $\rho_{EA}$ of the form (\ref{eq:cq-state2}). Thus it is enough to optimize over 
c-q states, which implies that the supremum is equal to unassisted capacity, 
due to (\ref{eq:clas-cap}). 
\end{proof}

To summarize, we have obtained, that for 
2EPR-GHZ box we have 
\be
0.3219 \lesssim CV\lesssim  0.469782.
\ee

\section{Conclusions}
\label{sec:conclusions}

In this paper we have considered the communication properties of entanglement redistribution tasks, e.g., Entanglement Swapping (ES), realized by means of LOCC protocols. By ``communication properties'' of said tasks, we mean,  on the one hand, the amount of classical communication that is needed to perform the redistribution -- its communication cost. On the other hand, we studied the communication value of the task, i.e., the amount of communication that is allowed by the bi- and multi-partite LOCC operation that realizes the redistribution. These two quantities may be thought as corresponding to two different points of view: that of providers and that of users, respectively. Providers are requested to realize the right LOCC operation: upon the correct input, i.e., the correct initial distribution of entanglement, such entanglement is redistributed ``as promised''. Users may instead feed whatever input to the redistribution operation, and in this way they may try to use the entanglement-redistribution ``machinery'', which is given to them by the providers, for a different aim than ``just'' entanglement redistribution, e.g., to communicate among themselves. On the one hand, the communication cost of the redistribution is a quantity of practical relevance; on the other hand, its communication value tells us whether the classical communication needed to implement the redistribution is, in a sense, irreversibly spent.

We showed that it is possible to estimate both the communication value and the communication cost irrespectively of the particular implementation of the redistribution process. Indeed, it is often useful to formulate information processing in terms of primitives rather than via specific realizations (e.g. a particular protocol).

The first step in studying the communication properties of entanglement redistribution has been a complete characterization of the structure of the LOCC operations involved. Using this characterization, we discovered a phenomenon of \emph{irreversibility} for ES: it needs 2 bits to be implemented and it can signal 1 bit. In contrast quantum teleportation needs 2 bits to be implemented and can also signal 2 bits. In this sense, ES is a weaker primitive than quantum teleportation. One can show that for $d$-dimensional system the gap between CC and CV for ES -- thus, between ES and teleportation -- is even greater. Namely, one needs $2\log d$ bits to implement ES, while in general it can be exploited to signal only 1 bit. As a primitive, ES may be split into two subprimitives, or intermediate steps (Figure \ref{fig:swappingsteps}): the transformation of two EPR pairs into a GHZ state, and of the latter into an EPR pair between any two subsystems.  We proved that both the CC and CV of the second (sub)primitive are equal to $1$, so that such primitive may be considered \emph{reversible} from the point of view of classical communication. With respect to the first subprimitive, there exists a box which has CC equal to $1$ and CV strictly less than $1$ and strictly greater than $0$. It would be interesting to find  the exact value of CV for this box.

We thank K. Horodecki for useful discussions. This work was supported by the European Commission through the Integrated Project FET/QIPC ``SCALA''. AG was also partially supported by Ministry of Science and Higher Education Grant No. N N206 2701 33.

\end{document}